\documentclass[aps,twocolumn]{revtex4}

\usepackage{epsfig}

\begin{document}

\title{Fourier's law from a chain of coupled anharmonic oscillators
under energy conserving noise}

\author{Gabriel T. Landi and M\'ario J. de Oliveira}

\affiliation{Instituto de F\'{\i}sica, Universidade de S\~ao Paulo,
Caixa Postal 66318, 05314-970, S\~ao Paulo, Brazil}

\begin{abstract}

We analyze the transport of heat along a chain
of particles interacting through anharmonic potentials
consisting of quartic terms in addition to harmonic quadratic
terms and subject to heat reservoirs at its ends. 
Each particle is also subject to an impulsive
shot noise with exponentially distributed waiting times whose effect 
is to change the sign of its velocity, thus conserving the energy
of the chain. We show that the introduction of this
energy conserving stochastic noise leads to Fourier's law. 
The behavior of thels heat conductivity for small intensities
of the shot noise and large system sizes are found to obey a 
finite-size scaling relation. We also show that the heat conductivity
is not constant but is an increasing monotonic function of temperature.

PACS numbers: 05.10.Gg, 05.70.Ln, 05.60.-k

\end{abstract}

\maketitle

\section{introduction}

The derivation of Fourier's law, or any other macroscopic law, 
from the microscopic underlying motion of particles constitutes
a major task in condensed matter physics.
This task comprises not only the derivation itself
but also the problem of setting up the appropriate microscopic model.
The simplest model one could conceive to derive Fourier's law
is a chain of particles interacting through harmonic potentials
in contact with two heat reservoirs at each end. 
However, it has been shown by Rieder et al 
\cite{rieder67}
that this model does not lead to Fourier's law.
Since then, several microscopic models have been introduced 
and studied
\cite{lepri97,dhar01,narayan02,grassberger02,casati03,deutsch03,
lepri03,eckmann04,pereira04,bonetto04,cipriani05,pereira06,delfini06,
mai07,pereira08,lukkarinen08,dhar08,basile09,iacobucci10,gersch10,dhar11}
some of them leading instead to the so called anomalous
Fourier's law.

Fourier's law states that the heat flux $J$ is proportional
to the gradient of the temperature $T$, that is,
\begin{equation}
J = - \kappa \frac{dT}{dx},
\end{equation}
where $\kappa$ is the heat conductivity. If we consider a
small bar of length $L$ subject to a difference in temperature
$\Delta T$, then $J=\kappa\Delta T/L$. Thus a microscopic model
for Fourier's law should predict a heat flux 
that decreases with $L$, for a fixed value of $\Delta T$,
according to 
\begin{equation}
J \sim \frac{1}{L}. 
\label{5}
\end{equation}
This amounts to saying that $\kappa$ is finite.
If, instead, we find that $J\sim L^{-\alpha}$ with $\alpha<1$, then
we are faced with the anomalous Fourier's Law.
In this case we may say that $\kappa$ is infinite,
as in the harmonic chain \cite{rieder67}, diverging
according to $\kappa\sim L^a$, with exponent $a=1-\alpha$.
Notice that $L$ should be macroscopically small, so that equation (\ref{5})
is the expression of Fourier's law, but microscopically
large so that a microscopic model for this law
should yield equation (\ref{5}) for sufficiently large $L$.

The heat flux $J$ of the linear harmonic chain 
placed between two heat reservoirs has been 
shown \cite{rieder67} to be independent of the size $L$ 
of the chain, meaning that the heat conductivity $\kappa$
is infinite. This result is a direct consequence 
of the ballistic transmission of heat by the elastic waves,
from one reservoir to the other.
A consequence of this result is that
a perfectly harmonic crystal has an infinite heat
conductivity \cite{ashcroft76}.
In real crystals the heat conductivity is manifestly finite.
This is due to the presence of lattice imperfections, impurities,
and other factors that act as scattering centers for the waves 
carrying energy, such as the Umklapp process \cite{peierls55}. 
These factors make the heat conduction a diffusion process
which implies Fourier's law.
The crossover from ballistic do diffusive behavior can also
be undertood in terms of the dephasing of the elastic waves
caused by the scattering events just mentioned. 
This process is characterized by a dephasing length \cite{dubi09a,dubi09b}
which, when smaller than the size of the system, gives the
linear temperature profile expected from Fourier's law.
Thus, a possible ingredient 
in the microscopic derivation of Fourier's law consists in
the presence of a diffusive motion at the microscopic level.

One way of introducing this ingredient is by means of stochastic
collisions that change the sign of the velocity of the particles.
This can be accomplished in the form of impulsive shot noises with
exponentially distributed waiting times \cite{dhar11}.
Two key properties are required when devising such a noise.
Firstly, it should conserve the total energy of the system
because any variation of the energy of the system should
only be due to the contact with the heat reservoirs.
Changing the sign of the velocity does not alter the energy.
Secondly, it should make the system ergodic even when
it is not coupled to the heat baths.
A harmonic chain with this type of shot noise has been indeed
studied by Dhar et al \cite{dhar11}. They showed that
this model can be mapped into the 
self consistent harmonic crystal 
introduced by Bolsteri et al \cite{bolsterli70},
and studied by Bonetto et al \cite{bonetto04}.
In this model each particle is in contact with independent heat
reservoirs, whose temperatures are chosen
so that, in the steady state, there is no exchange of heat
between these reservoirs and the chain. 
The contact with the reservoirs is regarded as a procedure to 
make the system ergodic \cite{bolsterli70}. This model
predicts a linear profile for the temperature and a finite
heat conductivity, which is independent of temperature.

Here we study a chain of particles interacting through anharmonic
forces in addition to random reversals of the velocity. More
specifically we consider a potential with quartic terms in
addition to the harmonic quadratic terms.
Without the stochastic shot noise, this is the well known
Fermi-Pasta-Ulam model in contact with heat reservoirs at
its ends, which was studied numerically by Lepri et al. \cite{lepri97}
who found a superdiffusive transport of heat implying an
anomalous Fourier's law. This important result has been confirmed by
other numerical studies 
and other approaches \cite{narayan02,lepri03,cipriani05,delfini06,
mai07,dhar08,lukkarinen08,beijeren12,roy12}
The impulsive stochastic shot noise that we use here
can be regarded as a procedure
that turns the superdiffusive transport of heat 
into a diffuse transport leading thus to Fourier's law. 
 
By numerically solving the Langevin equations for chains of
several sizes, we determine the heat conductivity as a
function of the system size $L$ and the rate of stochastic
collisions $\lambda$. When $\lambda$ is nonzero, we obtain
a finite heat conductivity and therefore Fourier's law.
For small values of $\lambda$, the heat conductivity is found to
behave as as $\kappa\sim\lambda^{-b}$, diverging when $\lambda=0$.
Our numerical results give $b=0.52\pm 0.06$. 
We have also determined the exponent
$a$ related to the divergence of $\kappa$ with $L$
at $\lambda=0$ and found $a=0.42\pm 0.04$.
These results are distinct from the harmonic case \cite{rieder67,dhar11} 
for which $a=1$ and $b=1$. 
Also, in contrast to the harmonic case, we have found
that the heat conductivity depends on
temperature. More precisely, for a fixed $\Delta T$,
we found that it increases monotonically
with the temperatures of the reservoirs.

\section{Model}

We consider a chain of $L$ interacting particles with equal masses $m$
and denote by $x_n$ the position of the $n$-th particle.
The total potential energy $V(x)=V(x_1,x_2,\ldots,x_L$) is considered to 
be a sum of anharmonic potential energies involving 
nearest-neighbor pairs of particles
\begin{equation}
V(x) = \sum_{n=0}^L \left[\frac{K_1}{2}(x_n-x_{n+1})^2 
+ \frac{K_2}{4}(x_n-x_{n+1})^4\right],
\end{equation}
where $K_1$ and $K_2$ are parameters. 
We consider fixed boundary conditions so that $x_{L+1}=x_0=0$.
When $K_2=0$, the harmonic potential is recast. The force $F_n$
acting on the $n$-th particle due to the potential $V(x)$ is
\[
F_n(x) = K_1(x_{n-1}+x_{n+1}-2x_n) +
\]
\begin{equation}
+K_2(x_{n-1}-x_n)^3 + K_2(x_{n+1}-x_n)^3 .
\end{equation}

\begin{figure}[t]
\centering
\epsfig{file=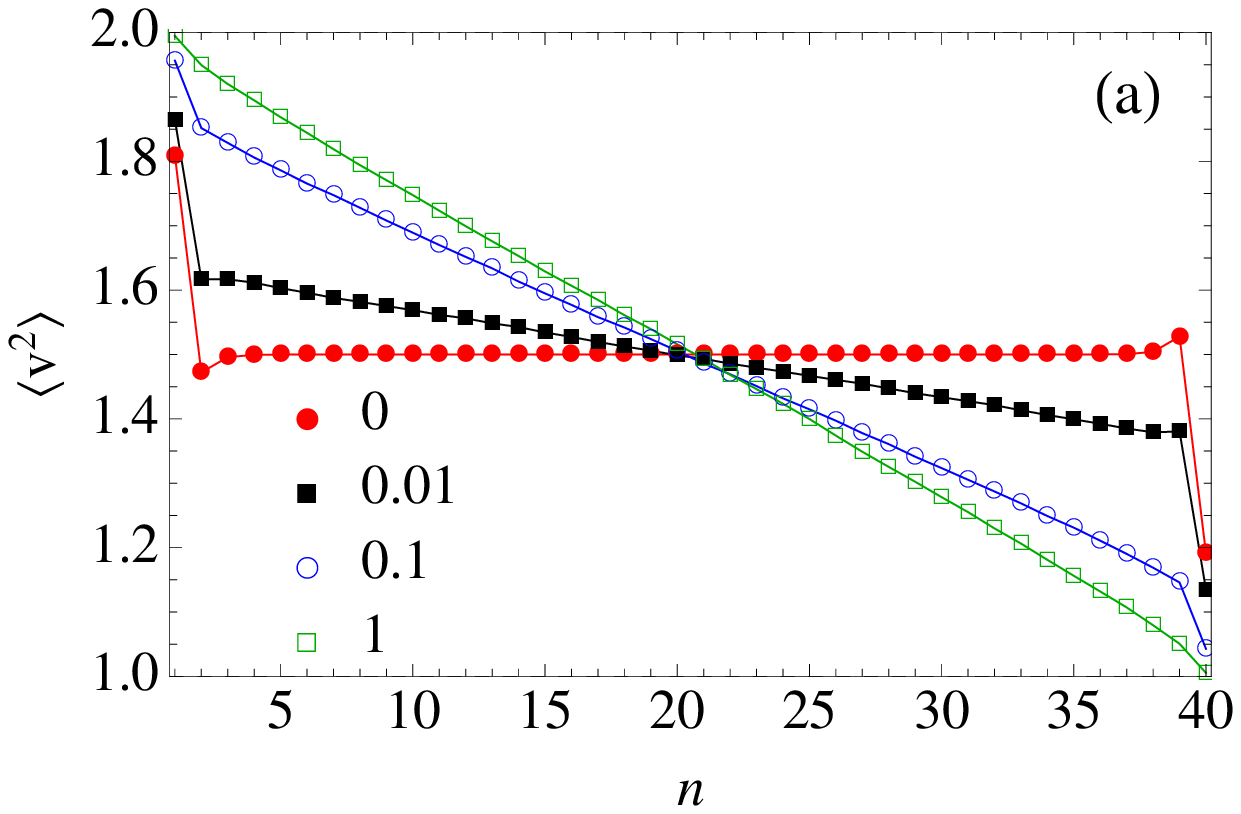,width=8.5cm}
\epsfig{file=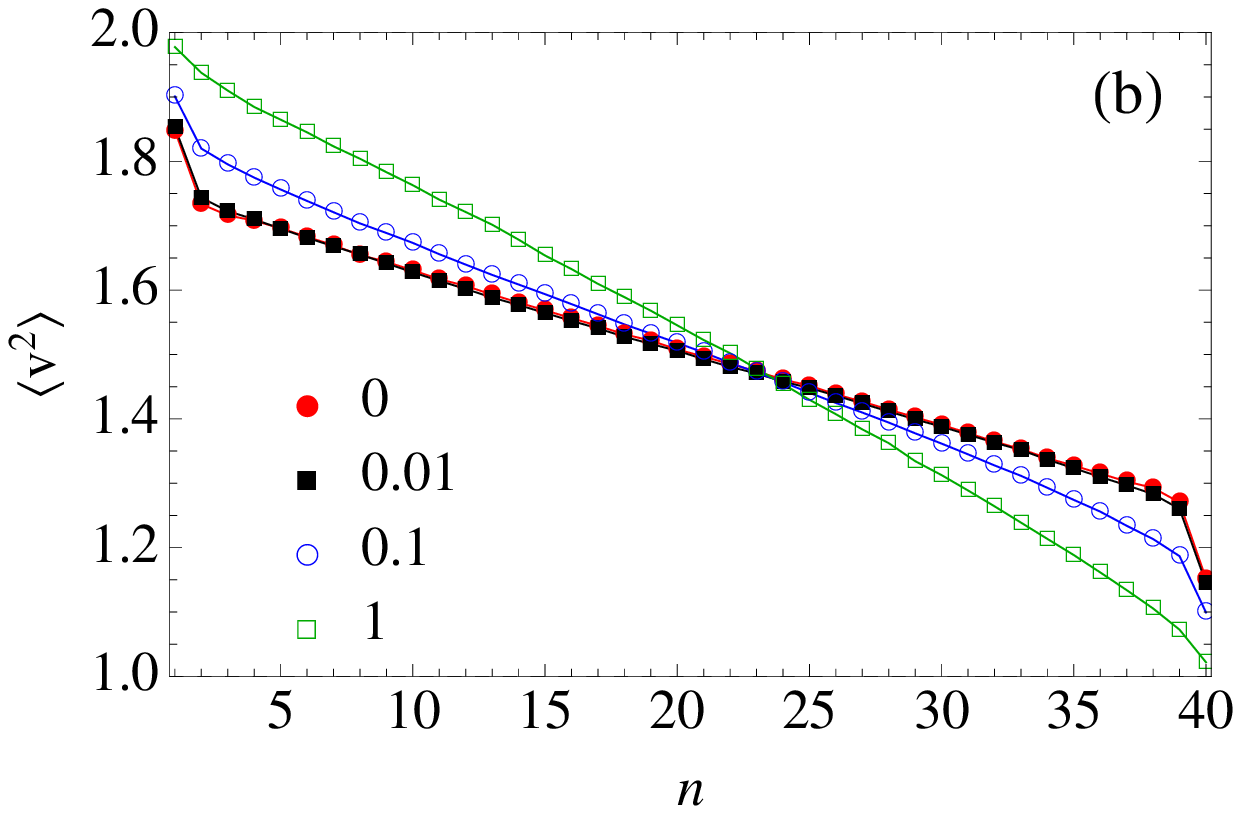,width=8.5cm}
\caption{Average squared velocity $\langle v_n^2\rangle$
of the particles as a function of the position $n$ on the chain,
which is in 
contact with heat reservoirs at temperatures $T_A=2$ and $T_B=1$.
Results are given for a chain of $L=40$ particles 
for the values of $\lambda$ indicated,
for the (a) harmonic case and (b) anharmonic case.
In (a) the continuous lines are exact results.}
\label{vel}
\end{figure}
 
The first and last particles
are connected to heat baths at temperatures $T_A$ and $T_B$,
and all particles are susceptive to stochastic collisions,
here described by forces ${\cal F}_n(t)$.
Denoting by $v_n=dx_n/dt$ the velocity of the $n$-th particle,
the equations of motion are stochastic equations given by
\begin{equation}
m \frac{dv_1}{dt} = F_1 + {\cal F}_1(t) 
- \alpha v_1 + \sqrt{2\alpha k_BT_A}\xi_A,
\label{1}  
\end{equation}
\begin{equation}
m \frac{dv_n}{dt} = F_n + {\cal F}_n(t),   
\qquad 2\leq n \leq L-1,
\label{2}
\end{equation}
\begin{equation}
m \frac{dv_L}{dt} = F_L + {\cal F}_L(t) 
- \alpha v_L + \sqrt{2\alpha k_BT_B}\xi_B,
\label{3}
\end{equation}
where $k_B$ is Boltzmann constant and
the last two terms in equations (\ref{1}) and (\ref{3})
represent the coupling to the heat reservoirs. The constant
$\alpha$ represents the strength of the coupling and 
$\xi_A$ and $\xi_B$ are independent Gaussian white noises with
zero mean and unit variance. The forces ${\cal F}_n(t)$ have the
form of impulsive shot noises, given by
\begin{equation}
{\cal F}_n = - 2m \sum_{\ell=1}^\infty v_n(t_{n\ell}^-)\delta(t-t_{n\ell}),
\end{equation}
where $t_{n\ell},\ell=1,2,\ldots$ are uncorrelated exponentially
distributed stochastic waiting times with a probability
density distribution $\rho(t) = \lambda e^{-\lambda t}$. Here, the 
parameter $\lambda$ is the rate of collisions, which has been taken
to be the same for all particles. After a collision occurring
at time $t$, the $n$-th particle changes its velocity
from $v_n(t^-)$ to $v_n(t^+)=-v_n(t^-)$, thus conserving
its kinetic energy and therefore the total energy
\begin{equation}
E = \sum_{n=1}^L\frac{m}2v_n^2 + V(x).
\label{12}
\end{equation}

Due to the contact with the reservoirs, 
the total energy $E$ is not a strictly conserved quantity.
From the stochastic equations of motion we find
\begin{equation}
\frac{dE}{dt} = {\cal J}_A + {\cal J}_B,
\end{equation}
where
\begin{equation}
{\cal J}_A = -\alpha v_1^2 + v_1\sqrt{2\alpha k_BT_A}\xi_A,
\label{13a}
\end{equation}
and
\begin{equation}
{\cal J}_B = -\alpha v_L^2 + v_L\sqrt{2\alpha k_BT_B}\xi_B.
\label{13b}
\end{equation}
In the stationary state, $\langle E\rangle$ is a constant
and the sum of the fluxes $J_A=\langle{\cal J}_A\rangle$ and
$J_B=\langle{\cal J}_B\rangle$ vanishes, that is, $J_B=-J_A$.
The heat flux $J=J_A$ can be calculated as the average of
the right-hand side of equation (\ref{13a})
or in a equivalent way from the equation
\begin{equation}
J = K_1\langle(x_{n-1}-x_n)v_n\rangle + 
K_2\langle(x_{n-1}-x_n)^3v_n\rangle,
\label{15}
\end{equation}
which we found to be numerically more accurate than the
formula $J=\langle{\cal J}_A\rangle$.

\begin{figure}[t]
\centering
\epsfig{file=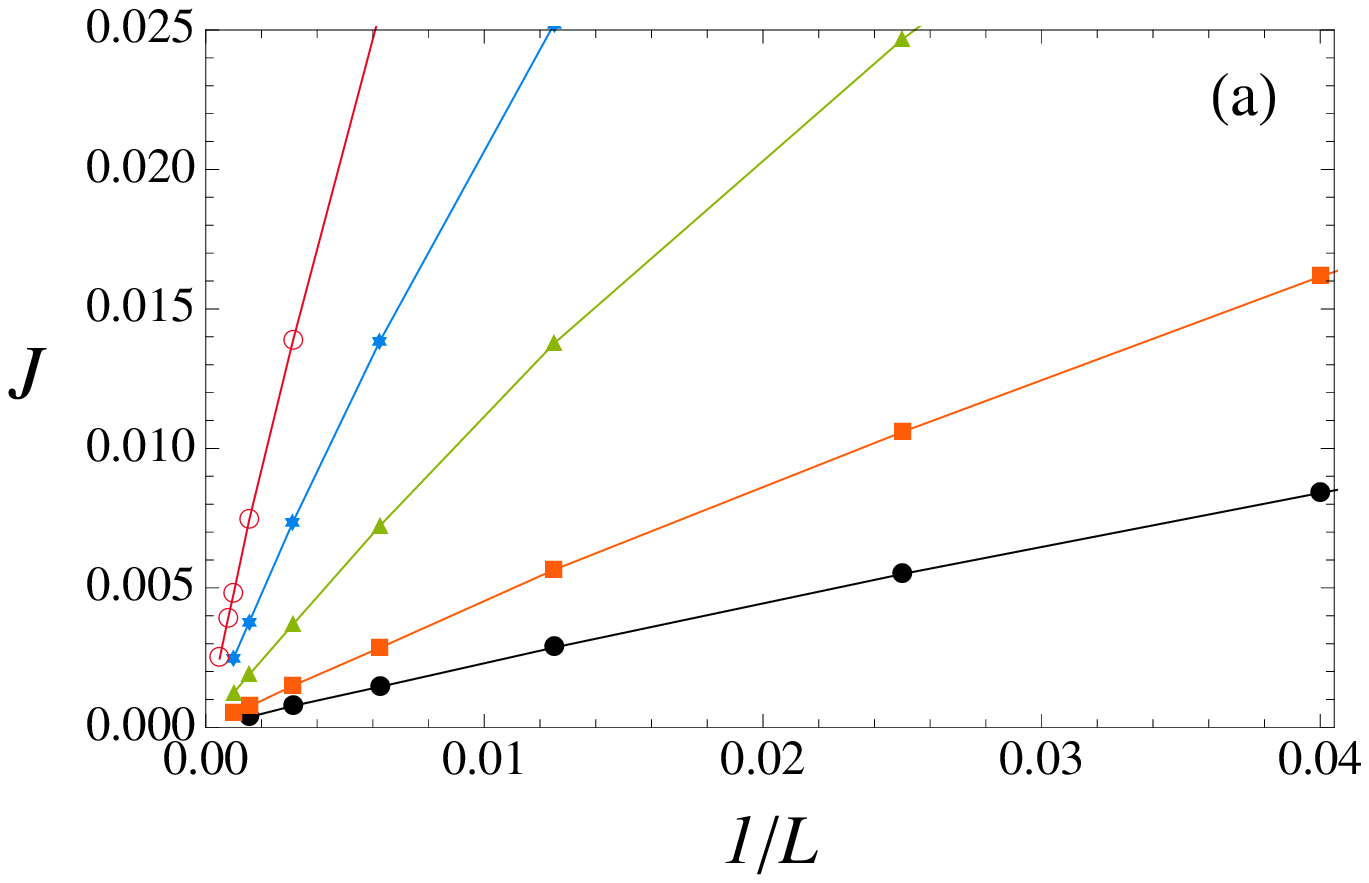,width=8.5cm}
\epsfig{file=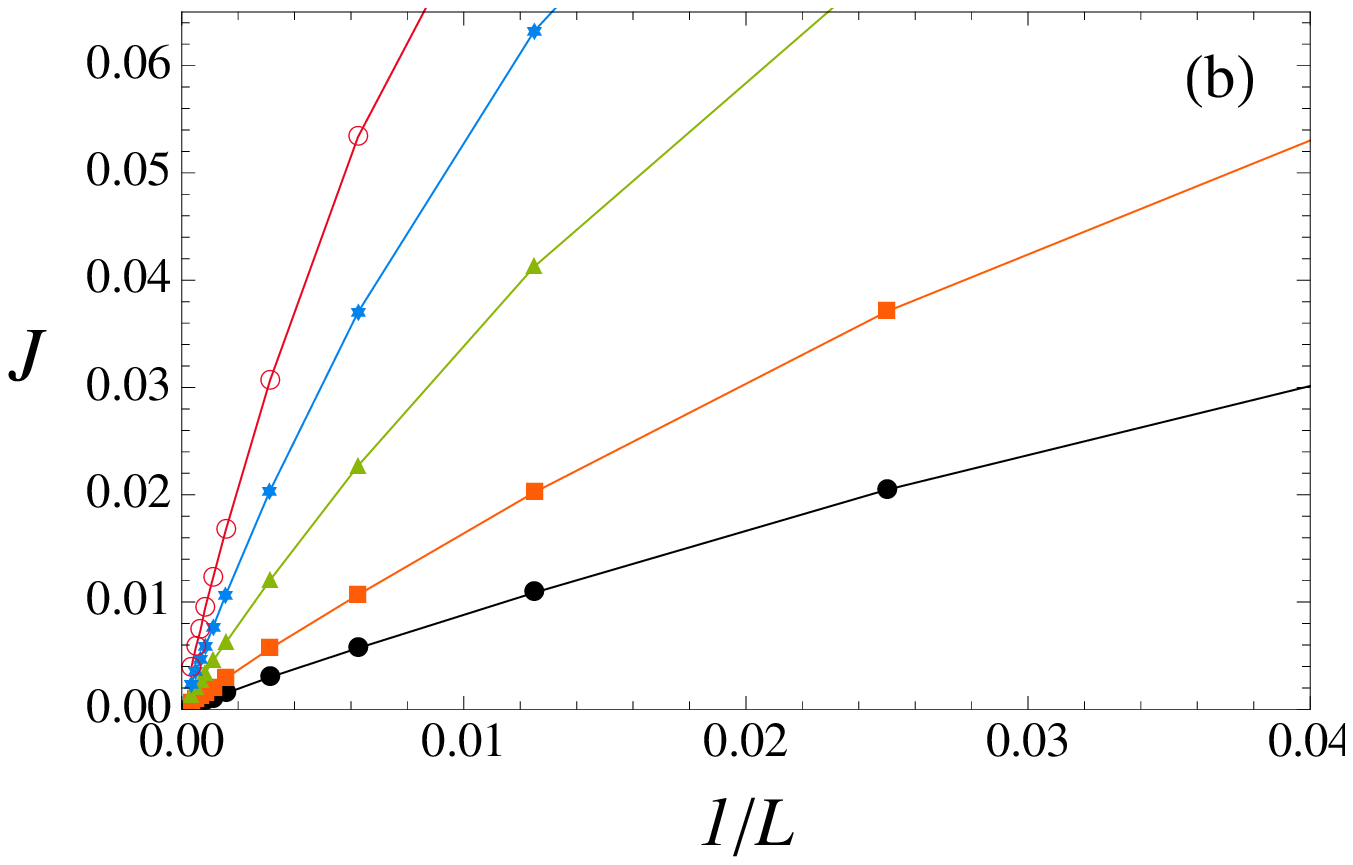,width=8.5cm}
\caption{Heat flux $J$ as a function of the inverse of
the system size $L$ for several values of $\lambda$
for (a) the harmonic chain and (b) the anharmonic chain.
In both cases, from top to bottom, $\lambda=0.05,0.1,0.2,0.5$ and $1$.
The temperatures of the reservoirs are $T_A=2$ and $T_B=1$.
In (a) the continuous lines are exact results.}
\label{flux}
\end{figure}

\section{Numerical solutions}

The stochastic equations of motion were solved numerically
for chains of several sizes $L$. This was done by discretizing
the time in intervals $\Delta t$. We use an approach in which
the deterministic part of the equations of motion of the inner
particles are handled
by the use of the Verlet algorithm \cite{verlet67} so as to ensure
that, in the absence of the heat baths, energy is conserved. For the
equations of motion of the first and last particles, which
contain the stochastic forces due to the heat baths, we used
the stochastic Verlet algorithm developed in \cite{paterlini98}.
As for the stochastic shot noises we treat them as follows. At each
time step, the sign of the velocity of each particle is 
changed with probability $p=\lambda\Delta t$. This procedure
generates a Poisson process with discrete waiting time $t=\ell\Delta t$
that is distributed according to the probability distribution
$p(1-p)^\ell$. In the continuous time limit, $\Delta t\to0$,
this yields the exponential distribution $\lambda e^{-\lambda t}$,
as required.

For definiteness, our numerical calculations were performed with
$k_B=1$, $m=1$, $\alpha=1$ and $\Delta t=0.01$.
For the anharmonic potential all results reported in this paper
were obtained for $K_1=1$ and $K_2=1$.
The size of the system ranged from $L=10$ up to $L=5000$.
We also present numerical results for the harmonic case ($K_2=0$), for
$K_1=1$ and compare with the results of the anharmonic case.
The existing exact solution \cite{dhar11} for harmonic case
is used to check our numerical procedure. 
The exact solution is obtained
by solving the equations for the pair correlations, which
is possible because they consist of closed equations.
However, this closure property does not happen for the
anharmonic case.

\begin{figure}[t]
\centering
\epsfig{file=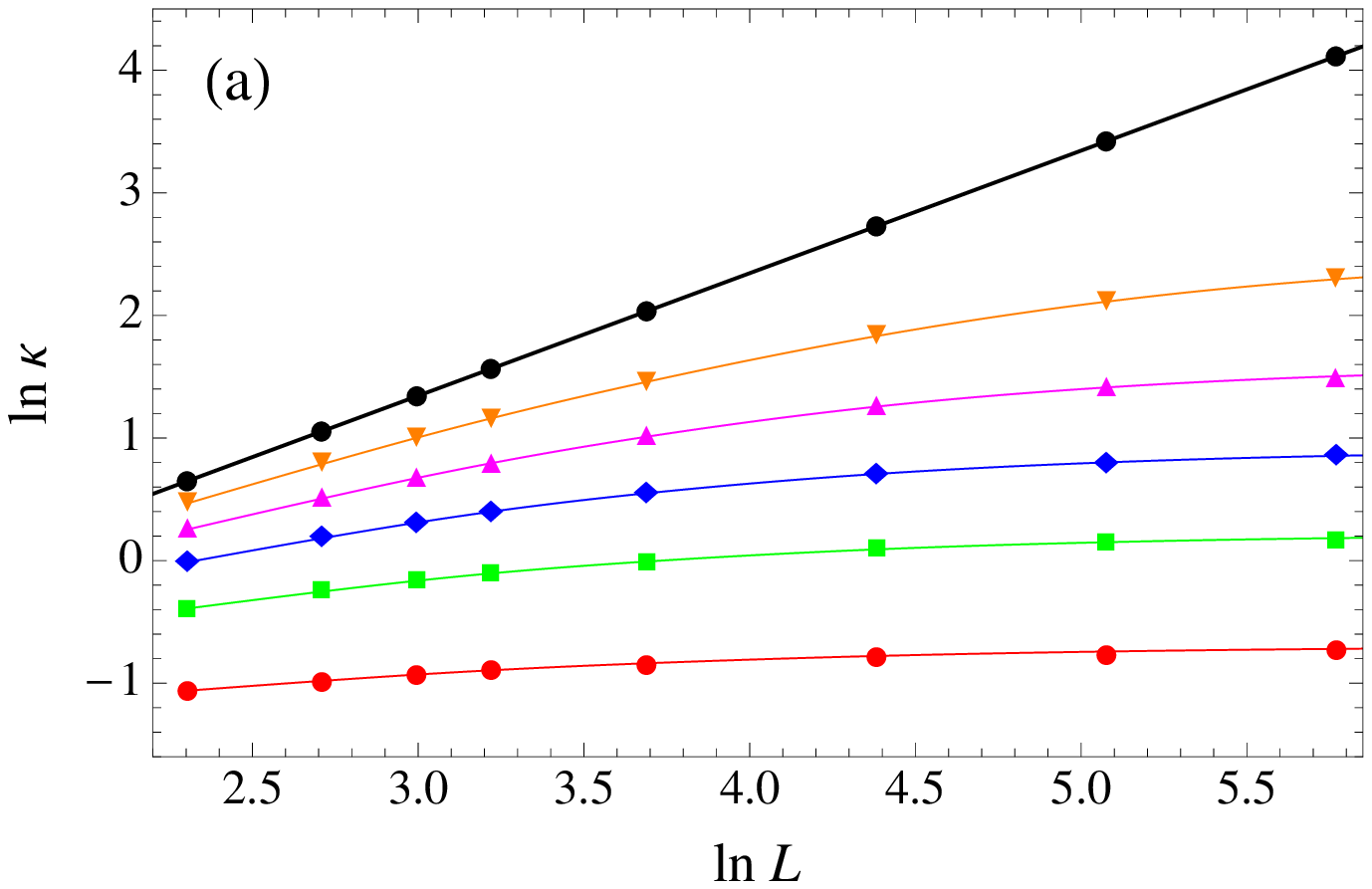,width=8cm}
\epsfig{file=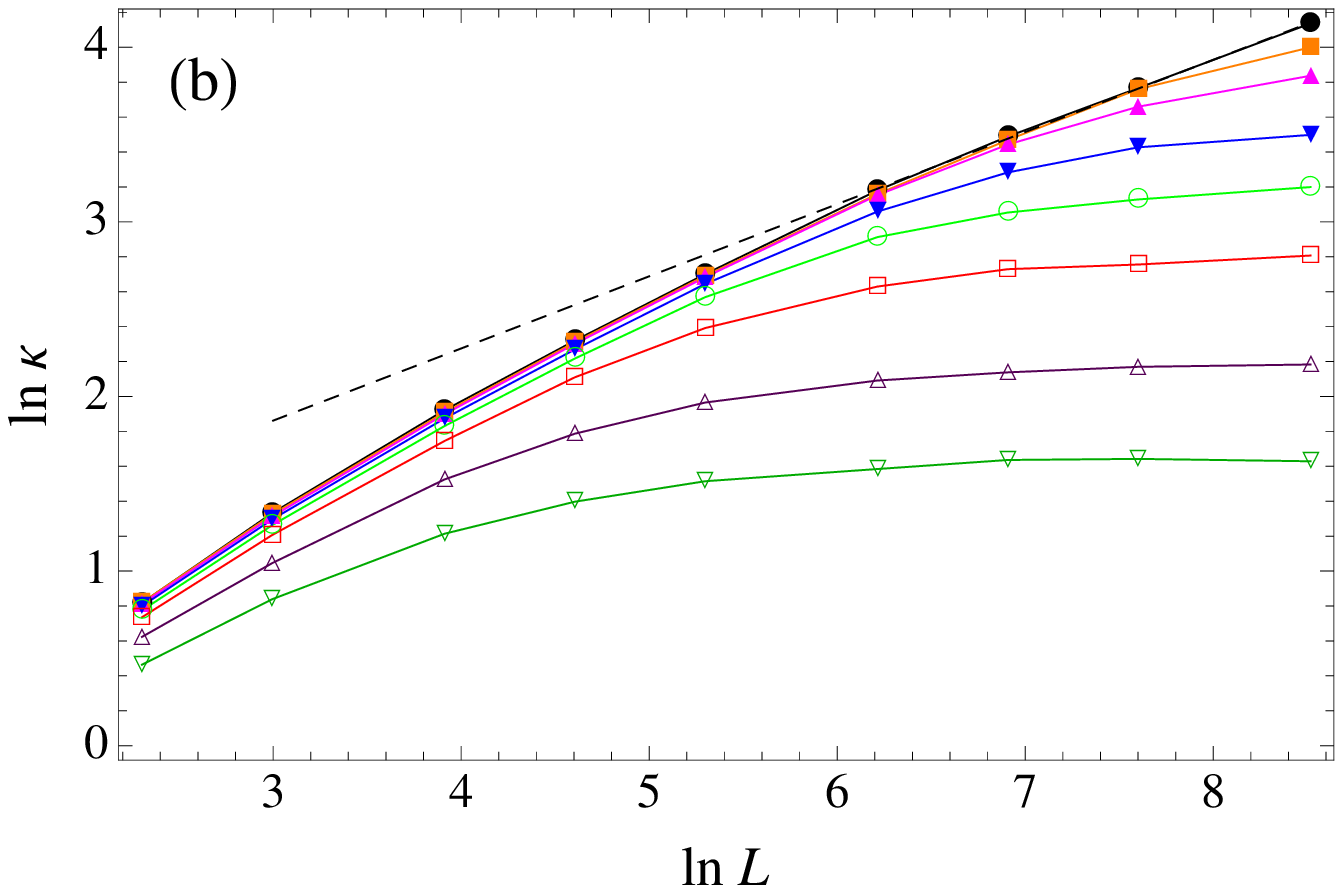,width=8cm}
\caption{Log-log plot of the heat conductivity $\kappa$
as a function of the system size $L$ for (a) the harmonic chain
and (b) the anharmonic chain. 
From top to bottom, $\lambda=0,0.02,0.05,0.1,0.2$ and $0.5$, for (a) 
and $\lambda=0,0.001,0.002,0.005,0.01,0.02,0.05$ and $0.1$, for (b).
The temperatures of the reservoirs are $T_A=2$ and $T_B=1$.
In (a), the continuous lines are exact results and
the slope of the straight line corresponding to $\lambda=0$ is $a=1$. 
In (b), the slope of the straight line (dashed line) 
fitted to the data points with large $L$ corresponding to $\lambda=0$
gives $a=0.42$.}
\label{kappa}
\end{figure}

In Fig. \ref{vel} we show the results for the
average kinetic energy for each particle as a function
of the position $n$ on the chain for the harmonic
and anharmonic cases. 
The temperatures of the reservoirs are considered to
be distinct, $T_A\neq T_B$, and our numerical calculations
were performed for $T_A=2$ and $T_B=1$.
Without stochastic collisions ($\lambda=0$)
the results of the harmonic case
shows that the kinetic energy is almost constant,
a result obtained by Rieder et al \cite {rieder67}, which
does not lead to Fourier's Law. 
The inclusion of the stochastic
collisions ($\lambda\neq0$) produces a drastically different
result. The average kinetic energy as a function of $n$
displays now a nonzero slope as can be seen in Fig. \ref{vel}.
For the anharmonic case, all curves,
including the case $\lambda=0$, show a nonzero slope.
In spite of that, the $\lambda=0$ case does not lead to Fourier's law.

We have calculated the flux $J$ at the stationary state
by using equation (\ref{15}) and the results are shown
in Fig. \ref{flux} as a function of $1/L$
for several values of $\lambda$.
From this figure we see clearly that $J\sim 1/L$,
for sufficiently large values of $L$, in
accordance with Fourier's law, as long as $\lambda\neq0$,
for both harmonic and anharmonic cases.
For sufficiently large values of $L$ 
the heat conductivity is given by $\kappa=JL/\Delta T$.
This quantity is plotted as a function of $L$
for several values of the rate of stochastic collisions $\lambda$,
including $\lambda=0$. The results are shown in Fig. \ref{kappa}.
For both the harmonic and anharmonic cases,
the heat conductivity $\kappa$ approaches a constant when $L\to\infty$,
as long as $\lambda\neq0$, and Fourier's law
is accomplished. When $\lambda=0$, our numerical results
gives a superdiffusive behavior with $\kappa\to\infty$
when $L\to\infty$, 
according to
\begin{equation}
\kappa \sim L^a, \qquad\qquad \lambda=0,
\end{equation}
as shown in Fig. \ref{kappa}.
For the harmonic case, $a=1$,
which is in accordance with
the result by Rieder et al \cite{rieder67}.
For the anharmonic case we get $a=0.42\pm 0.04$.
This is in agreement with the result $a=0.45\pm 5$ found by 
Lepri et al. \cite{lepri97} and in excelent agreement
with the value $a=2/5$ \cite{lepri03,lukkarinen08}.

\begin{figure}[t]
\centering
\epsfig{file=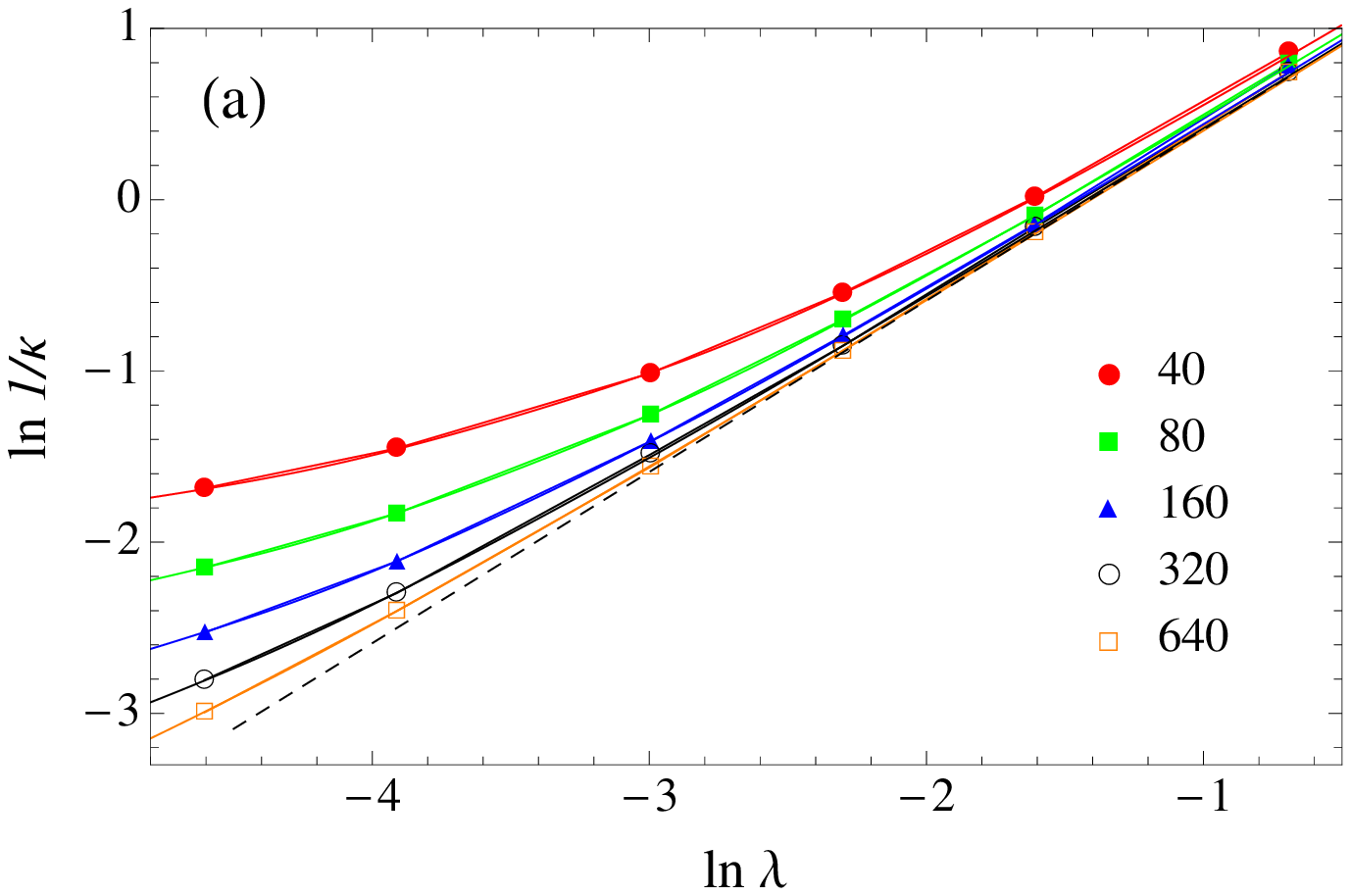,width=8.5cm}
\epsfig{file=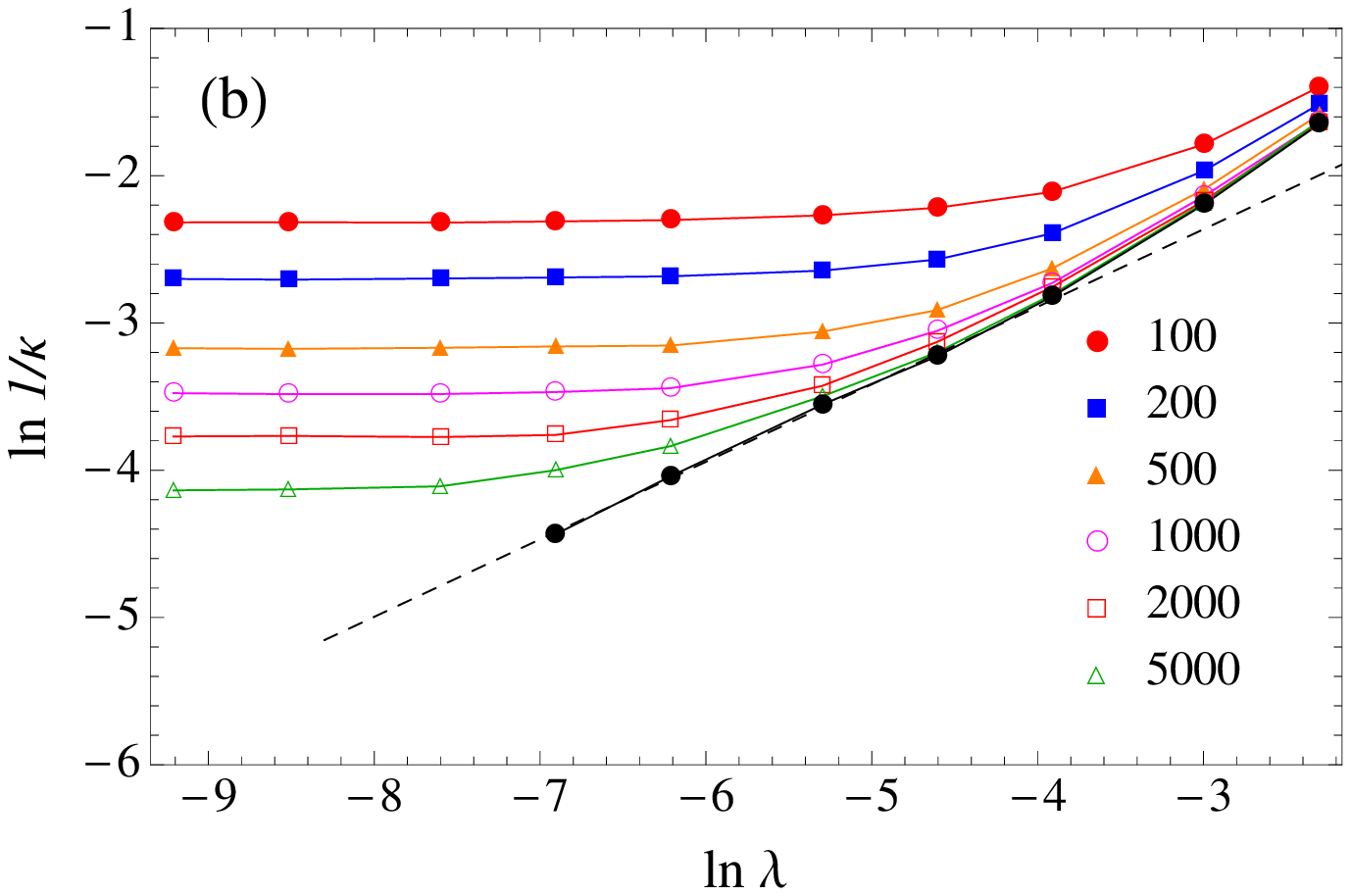,width=8.5cm}
\caption{Log-log plot of the reciprocal of the heat conductivity $\kappa$
as a function of the collision noise $\lambda$
for (a) the harmonic chain and (b) the anharmonic chain,
for several values of $L$ indicated.
The temperatures of the reservoirs are $T_A=2$ and $T_B=1$.
In (a) the continuous lines are exact results.
The slope of the straight dashed lines 
is $b=1$ for (a) the harmonic case and 
$b=0.52$ for the (b) anharmonic case.
The lowest curve in both cases are extrapolation
obtained from finite values of $L$.}
\label{kappalambda}
\end{figure}

To analyze the behavior of the heat conductivity
$\kappa=JL/\Delta T$ as $\lambda\to0$ we have plotted
this quantity as a function of $\lambda$ for several
values of the system size $L$.
The results are shown in Fig. \ref{kappalambda}.
We have plotted also the extrapolated
values of $\kappa$ when $L\to\infty$ for each $\lambda$. 
These extrapolated values were extracted from the 
plot of $1/\kappa$ versus $1/L$. The heat conductivity
$\kappa$ diverges when $\lambda\to0$ according to
\begin{equation}
\kappa \sim \lambda^{-b}, \qquad\qquad L\to\infty.
\end{equation}
For the harmonic case our results give $b=1$,
a result obtained by Dhar et al \cite{dhar11}.
For the anharmonic case we found $b=0.52\pm 0.06$,
a result clearly distinct from the harmonic case.

\begin{figure}[t]
\centering
\epsfig{file=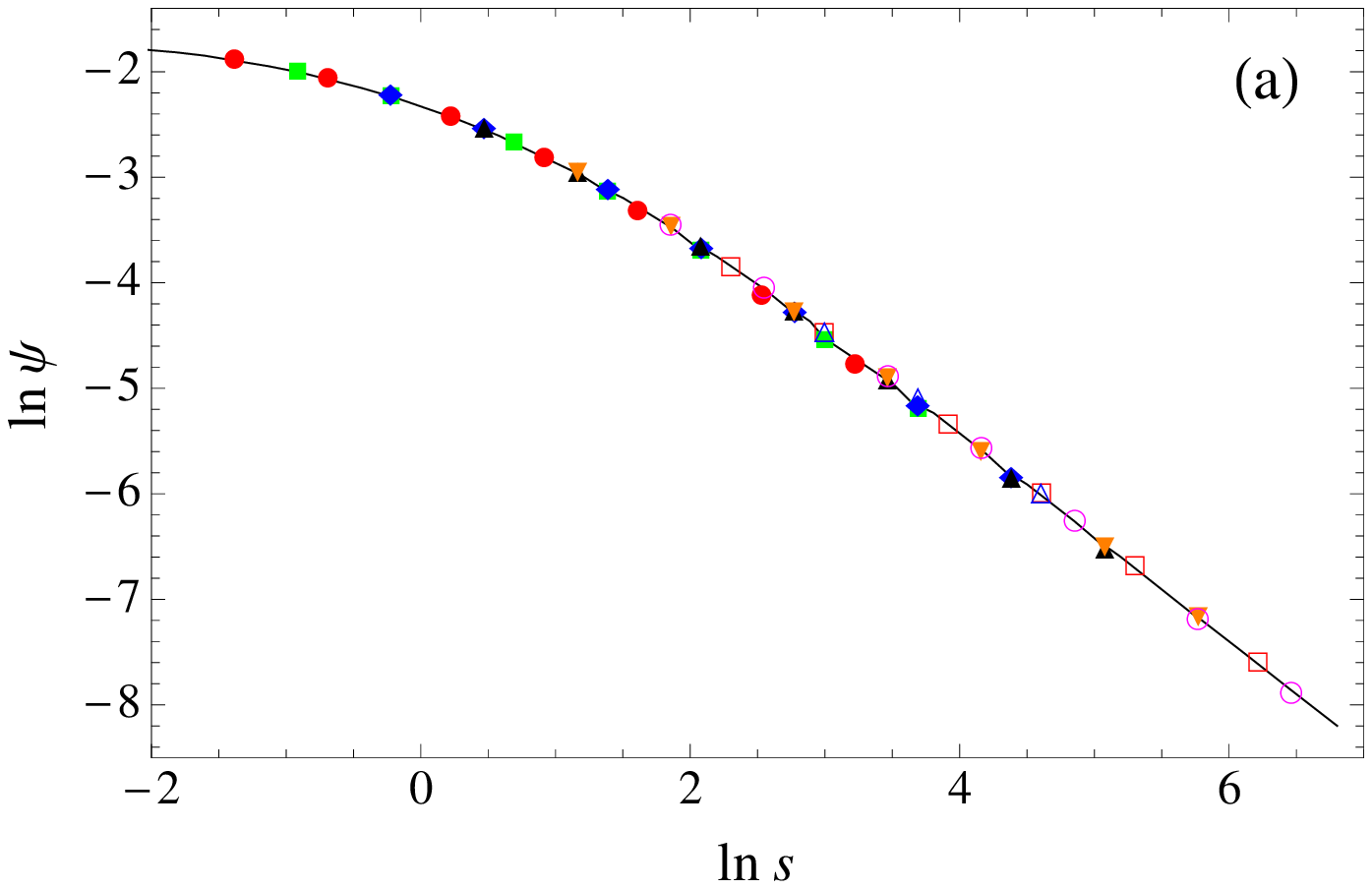,width=8cm}
\epsfig{file=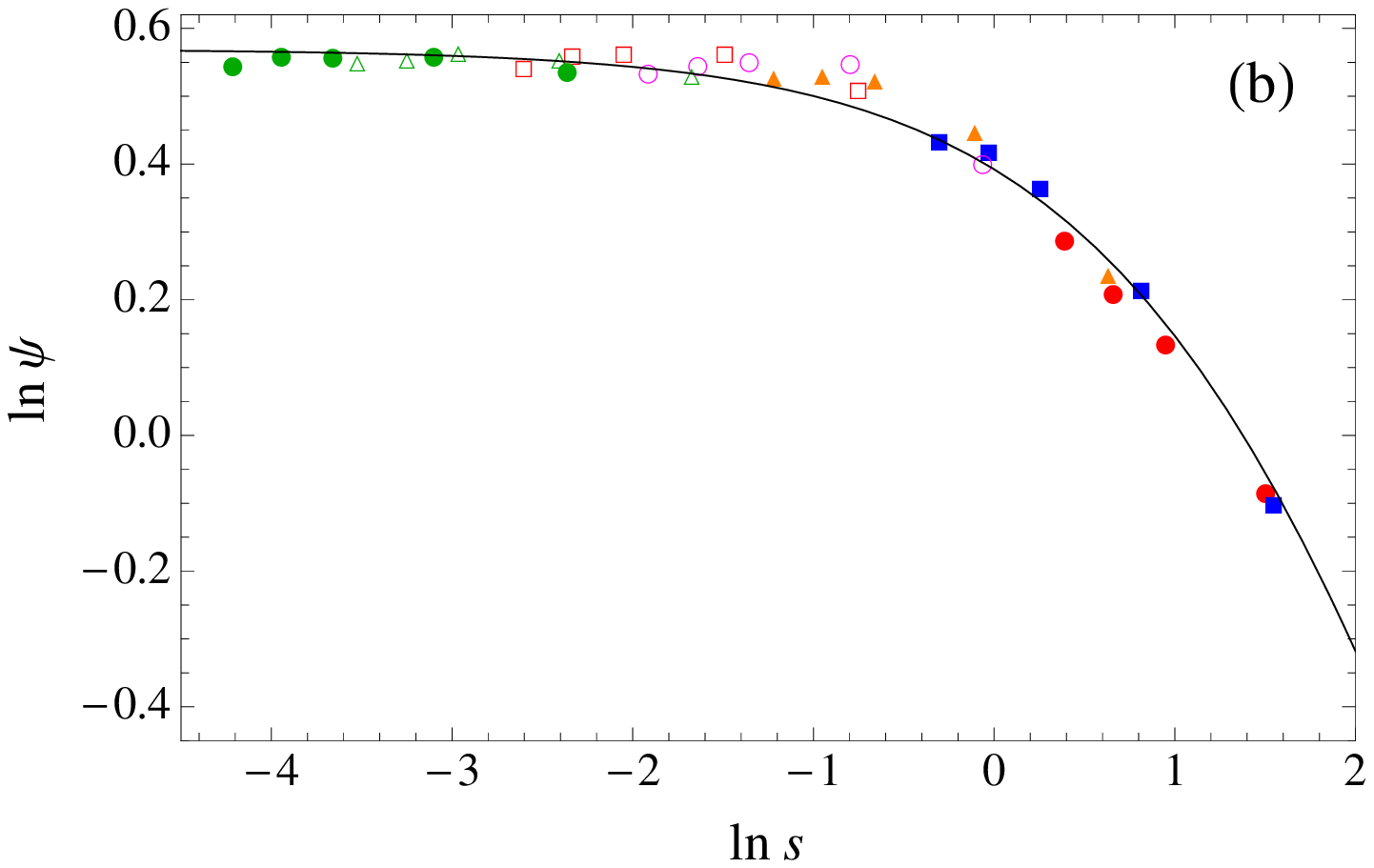,width=8cm}
\caption{Data colapse obtained from the numerical results
of the heat conductivity $\kappa$ obtained from several
values of $\lambda$ and $L$ corresponding to the harmonic case
(a) and the anharmonic case (b),
where $\Psi=\kappa/L^a$ and $s=\lambda L^{a/b}$. 
The continuous line in case (a) is described by the expression 
(\ref{17}) with $A=1/4$ and $B=3/2$.
In case (b), the continous line is a guide to the eyes. 
The temperatures of the reservoirs are $T_A=2$ and $T_B=1$.
Points with the same symbol correspond to the same value
of $\lambda$.}
\label{colaps}
\end{figure}

The algebraic behavior of $\kappa$ with $L$ and $\lambda$
can be obtained by assuming the following finite-size
scaling for the heat conductivity,
\begin{equation}
\kappa = L^a \psi(\lambda L^c),
\label{23}
\end{equation}
where $\psi(s)$ is a universal function of $s=\lambda L^c$
such that $\psi(0)$ is a finite constant and
$\psi(s)\sim s^{-b}$ when $s$ is large. 
To ensure a finite conductivity 
in the limt $L\to\infty$, the exponent $c$ must be
related to the exponents $a$ and $b$ by $c=a/b$. 
In Fig. \ref{colaps}, we show the data collapse for the 
harmonic case by plotting
$\kappa/L^a$ as a function of $\lambda L^c$, where in this case
$a=1$ and $c=1$. The data collapse is well described by the expression
\begin{equation}
\psi(s) = \frac{A}{B+s},
\label{17}
\end{equation}
as can be seen in Fig. \ref{colaps}.
This function was obtained by numerically solving the exact
equations for the pair correlations from which we found
$A=1/4$ and $B=3/2$.
For large values of $s$, this result gives $\psi\sim s^{-1}$
so that $\kappa\sim \lambda^{-1}$, independent of $L$,
which is the behavior of the conductivity for the harmonic 
chain when $L\to\infty$ \cite{bolsterli70,bonetto04,dhar11}.
For the anharmonic case, the data colapse, shown in 
Fig. \ref{colaps}, was obtained by using the exponents
$a$ and $b$ obtained previously.

It is worth mentioning that Lepri et al. \cite{lepri09}
in their study of a one-dimensional harmonic crystal
with energy conservative noise consisting of elastic collisions between
neighboring particles reported a conductivity 
that behaves as $(L/\lambda)^{1/2}$ for large values of $L$.
Assuming for this system a scaling function of the form (\ref{23})
and keeping in mind that $a=1$, because 
the conductivity behaves as $\kappa\sim L$
when $\lambda=0$, we may conclude that 
$b=1/2$ and $c=1$. Notice, however, that $c\neq a/b$ so that
$\kappa$ is not finite but diverges when $L\to\infty$.

A relevant feature of the present anharmonic chain
with random reversal of velocities is 
the dependence of the heat conductivity with temperature.
For the harmonic case the heat conductivity
is temperature independent \cite{bolsterli70,dhar11},
a result that can be understood by using the following reasoning.
If we rescale the temperature of the reservoirs by
a factor $r$, that is, $T_A\to rT_A$ and $T_B\to rT_B$,
and the positions and velocities by a factor $r^{1/2}$,
that is, $v_n\to r^{1/2}v_n$ and $x_n\to r^{1/2}x_n$,
the equations of motions for the case $K_2=0$
become invariant. In addition,
according to Eq. (\ref{15}) with $K_2=0$,
the heat flux changes as the temperature, that is,
$J\to r J$. From this relation it follows that $J$ is
a homogeneous function of $T_A$ and $T_B$, so that
$J=T_A\phi(T_B/T_A)$. Writing $T_B=T_A+\Delta T$, then
for small values of $\Delta T$ it follows that 
$J\sim \Delta T$ leading to a heat conductivity $\kappa$
independent of temperature, a result that we have also checked
numerically.

\begin{figure}[t]
\centering
\epsfig{file=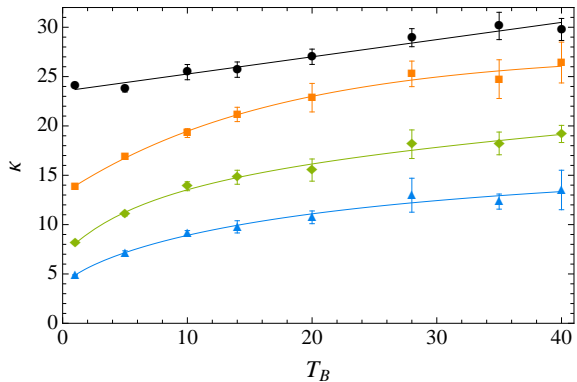,width=8cm}
\caption{Heat conductivity $\kappa=JL/\Delta T$ as a function of the
temperature $T=T_B$ of the colder reservoir,
for several values of $\lambda$. From top to bottom
$\lambda=0,0.02,0.05$ and $0.1$.
All results were obtained for a chain of size $L=500$ and
by using the same value of the difference
in temperature of the reservoirs, $\Delta T=1$.
The continuous lines are guide to the eyes.}
\label{deptemp}
\end{figure}

For the anharmonic case we have found that the 
heat conductivity $\kappa$ is an increasing function of temperature.
In Fig. \ref{deptemp} we show $\kappa$ as a function
of the temperature $T=T_B$ of the colder reservoir,
The heat conductivity was determined
from $\kappa=JL/\Delta T$, for the same value of
$\Delta T=1$ and for several values of the noise
parameter $\lambda$. We used $L=500$, a value high enough
so that the values of $\kappa$ may be considered the 
asympototic values (see Fig. \ref{kappa}),
with the exception of the case $\lambda=0$.
Our results indicate a monotonic increase of the heat
conductivity with temperature as can be
seen in Fig. \ref{deptemp}.
We have found that our results are consistent with 
the result of Aoki and Kusnezov \cite{aoki01}
and with the upper bounds of Bernardin and Olla
\cite{bernardin11}.

\section{Conclusion}

In conclusion, we have considered a chain of particles interacting
through anharmonic potentials and subject to heat reservoirs at
its ends. In addition, the chain is subject to a shot noise
that changes the sign of the velocities of the particles at random times,
distributed according to an exponential distribution.
The shot noise does not change the energy so that the
changes in energy are only due to the contact with the
reservoirs. We have shown that, 
when the chain is connected to reservoirs
at different temperatures, the heat flux
is inversely proportional to the size of the system,
as long as the shot noise parameter $\lambda$ is nonzero, and therefore
in accordance with Fourier's law. 
Our results suggest, in accordance with \cite{bolsterli70},
that the ergodicity may play a crucial
role in the derivation of Fourier's law.

We have also obtained the behavior of $\kappa$ with $L$ 
when $\lambda=0$ and $\kappa$ with $\lambda$ when $L\to\infty$.
Both behaviors are found to be algebraic, characterized
by exponents $a=0.42\pm0.04$ and $b=0.52\pm0.06$.
This allows us to introduce a finite-size scaling
in which the noise parameter
scales with the inverse of the system size to the power
$a=b/c$. For the harmonic case, we have shown that
a finite size scaling exists, a result that has been also
obtained by numerically solving the equations
for the correlations.

\section*{Acknowledgment}

We wish to thank E. Pereira for his helpful comments.
We acknowledge the Brazilian agencies FAPESP and CNPq
for financial support.


\end{document}